\documentclass[showpacs,preprintnumbers,amsmath,floatfix]{revtex4}
\usepackage{graphicx}
\usepackage{amssymb}
\usepackage{amsmath}
\usepackage{float}
\usepackage{caption}
\usepackage{rotating}
\usepackage[normalem]{ulem} 
\begin{document}

\title{The Dynamics of EEG Entropy}

\author{M. Ignaccolo$^{1}$, M. Latka$^{2}$, W. Jernajczyk$^{3}$, P. Grigolini$^{4}$
and B.J. West$^{1,5}$ \\
%EndAName
1) Physics Department, Duke University, Durham, NC\\
2) Institute of Biomedical Engineering, Wroclaw University of Technology,
Wroclaw, Poland \\
3) Department of Clinical Neurophysiology, Institute of Psychiatry and
Neurology, Warsaw, Poland \\
4) Center for Nonlinear Science, University of North Texas, Denton, TX\\
5) Mathematics and Information Science Directorate, Army Research Office}

\date{\today}

\begin{abstract}
EEG time series are analyzed using the diffusion entropy method. The
resulting EEG entropy manifests short-time scaling, asymptotic saturation
and an attenuated alpha-rhythm modulation. These properties are faithfully
modeled by a phenomenological Langevin equation interpreted within a neural
network context.
%\keywords{EEG \and  Entropy \and Statistical Analysis} 
%\PACS{87.19.le \and  05.40.-a \and  05.10.Gg \and 87.19.L-}
%\PACS{First \and Second \and More}
\end{abstract}

\maketitle

\section{Introduction}
In the nearly one hundred years since the electroencephalogram (EEG) was
introduced into neuroscience there have been a variety of methods used in
attempts to establish a taxonomy of EEG patterns in order to delineate the
correspondence between brain wave patterns and brain activity. Over that
time the single channel EEG time series has been interpreted as consisting
of relatively slow regular variations called \textit{signal} and the
relatively rapid erratic fluctuations called \textit{noise}. This separation
implies that the signal contains information about the EEG channels in the
brain, whereas the erratic fluctuations are a property of a channel's
environment and does not contain any useful information. Recent studies have
refined this engineering model and extracted information from the random
fluctuations by concentrating on what is believed to be the scaling behavior
of the time series. Analysis of the second moment of the single channel time
series shows an algebraic scaling in time, i.e., $\left\langle
X(t)^{2}\right\rangle \varpropto t^{2H}.$ The brackets denote a suitably
defined averaging procedure and detrended fluctuation analysis (DFA) \cite
{dfa1}, which measures the standard deviation of the detrended fluctuations,
has been the method of choice for the recent analysis of EEG data \cite
{scaling}. Buiatti et al.~\cite{buiatti} use DFA to show that specific
task-demands can modify the temporal scale-free dynamics of the ongoing
brain activity as measured by the scaling index.

The 'signal' parts of the EEG time series are called waves or rhythms. The
nature and scope of these waves have been widely investigated, see Ba\c{s}ar 
\cite{basar06} for a review. The alpha rhythm (7-12 Hz) has been shown to be
typical of awake individuals under no stimulation. Ba\c{s}ar, along with
colleagues, have developed an integrative theory of alpha oscillations in
brain functioning \cite{basar97}. They hypothesize that there is not one,
but several alpha generators distributed within the brain and note that the
alpha rhythm may act as a nonlinear clock in the manner suggested by Wiener 
\cite{wiener58} to serve as a gating function to facilitate the association
mechanisms in the brain.

As a measure of order/disorder, entropy has been used to characterize EEG
signals. Schl\"{o}gl et al.~\cite{schogl99} measured the entropy of 16 bits
EEG polysomnograhic records and found it in the range of 8-11 bits. Inouye
et al.~\cite{inouye91} employed spectral entropy, as defined by the Fourier
power spectrum, but the fact that EEG time series are not stationary, in the
sense that the autocorrelation function is not simply a function of the
two-time difference, obviates the use of Fourier transforms. Subsequently,
wavelet entropy was used by Rosso et al.~\cite{rosso01} to study the
order/disorder dynamics in short duration EEG signals including evoked
response potentials. Patel et al.~\cite{patel99}, using a combination of MRI
and entropy maximization, demonstrated that the generators of alpha rhythm
are mainly concentrated over the posterior regions of the cortex.

The diffusion entropy (DE) method \cite{dea} has been successfully used to
discriminate between the contributions of the low frequency waves (signal)
and the high frequency fluctuations (noise), e.g., the influence of the
seasons on the daily number of teen births in Texas \cite{bimbini} and the
effect of solar cycles on the statistics of solar flares \cite{flares}. In
this Letter we use the DE method to provide insight on the low/high
frequency dynamics of EEG time series.

\section{EEG analysis}
Each single channel recording of the EEG time series consists of a sequence
of $N+1$ data points, and the difference between successive data points is
denoted by $\xi _{j}$ for $j=1,2,..,N$. For the DE analysis a set of
stochastic variables $X_{k}(t)$ is constructed from the differenced data
points in the following way 
\begin{equation}
X_{k}\left( t\right) =\overset{k+t}{\underset{j=k}{\sum }}\xi _{j},\text{ }%
k=1,2,..,N-t+1  \label{rndwlk}
\end{equation}
to obtain $M=N-t+1$ replicas of a stochastic trajectory by means of
overlapping windows of length $t$. This ensemble of trajectories, generated
by the EEG time series, is used to construct the histogram using the number
of trajectories falling in a specified interval to estimate the \textit{pdf} 
$p(x,t).$ The pdf is then used to calculated the information entropy, a
quantity introduced in discrete form for coding information by Shannon \cite
{shannon48} and in continuous form for studying the problem of noise and
messages in electrical filters by Wiener \cite{wiener48}. We use the latter
form here, 
\begin{equation}
S(t)=-\int p(x,t)\log _{2}p(x,t)dx.  \label{entropy}
\end{equation}
The entropy $S(t)$ of Eq.~(\ref{entropy}) assumes a simple analytical form
if the pdf of the diffusion process satisfies the scaling relation: 
\begin{equation}
p(x,t)=\frac{1}{\sigma \left( t\right) }F\left( \frac{x}{\sigma \left(
t\right) }\right) \Rightarrow S(t)=C+\log _{2}\sigma (t)  \label{scale}
\end{equation}
where $C$$=$$-\int F(y)\log _{2}F(y)dy$ is a constant and $\sigma \left(
t\right) $ for a gaussian diffusion process is the time-dependent standard
deviation $\sigma (t)$. More generally an $\alpha $$-$ stable L\'{e}vy
process also scales in this way, in which case $\sigma (t)$ is more general
than the standard deviation of the underlying process.

If the time series were scaling, as assumed in a number of analyses of EEG
data \cite{scaling}, the 'variance' would be $\sigma $$(t)$$\varpropto $$%
t^{\delta }$ and the entropy graphed versus time on log-linear graph paper
would increase linearly with slope $\delta $. Consequently, the way in which
the entropy for a time series scales is indicative of the scaling behavior
of the time series. Note that in a simple diffusive process this index is
equal to the one obtained from the second moment, that is, $\delta $$=$$H$.
However, in general, even when there is scaling $\delta $$\neq $$H$ and in
the case of EEG time series we establish that there is no scaling at all.

We now consider EEG signals of 20 awake individuals in the absence of
external stimulations (quiet, closed eyes). EEG signals were recorded using
the 10-20 international recording scheme. For 8 individuals only the
channels O1,O2,C3 and C4 were recorded, for the remaining 12 individuals all
the channels are available. To have a coherent database, we restrict our
analysis to the channels O1,O2,C3 and C4, which are the channels
traditionally used in sleep studies. The sampling frequency of all EEG
records is 250Hz. Durations of EEG records vary from 55s to 400s with an
average duration of 128s.

\begin{figure}[tbp]
\includegraphics[angle=-90,width=1.0\linewidth]{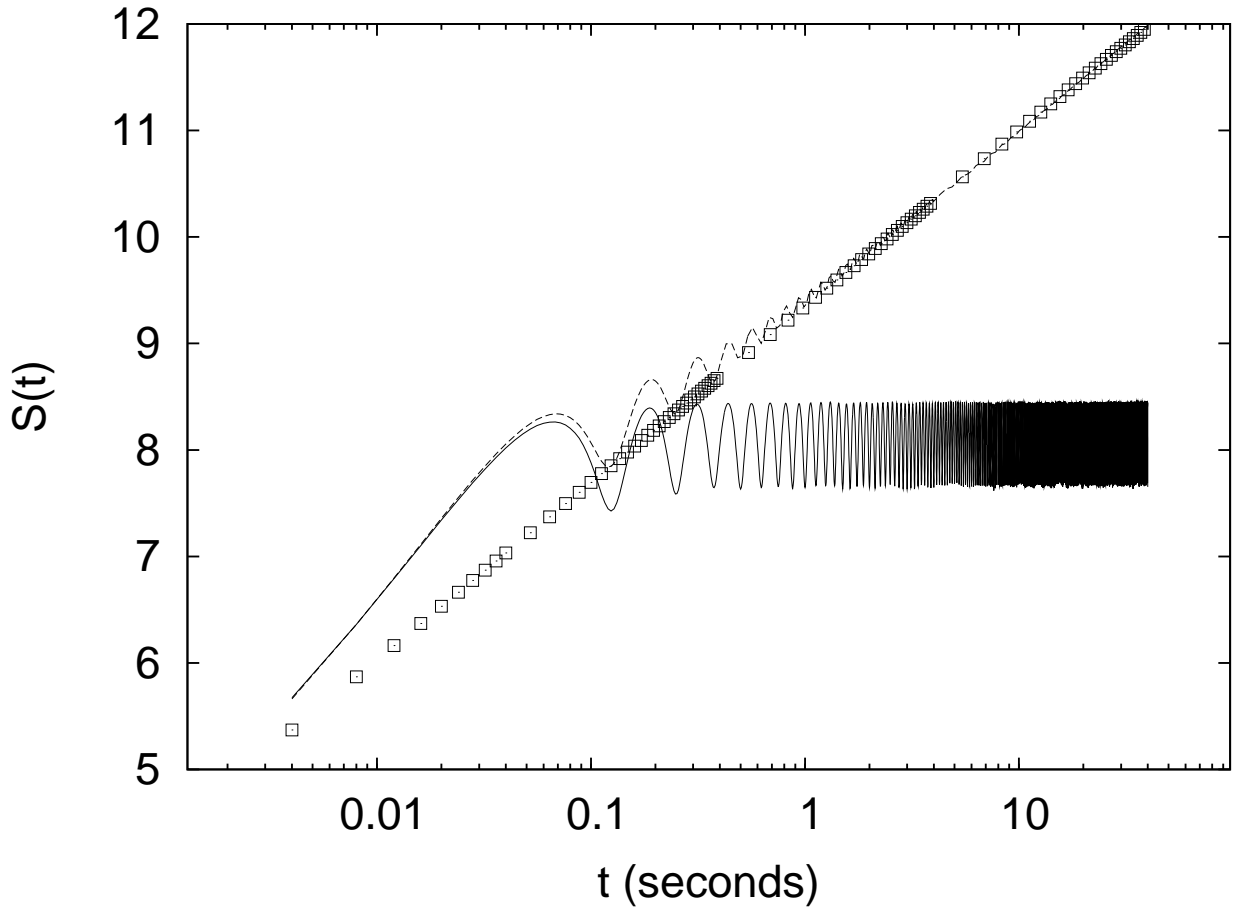}
\caption{The diffusion entropy $S(t)$ calculated using the increments of the
channels O1, O2, C3 and C4 for one of the 20 subjects considered in this
study. The inset depicts the pdfs $\text{p}_{\text{sat}}(x)$$=$$p(x,t=2000)$
for each channel: squares (O1), circles (O2), upward triangles (C3), and
downward triangles (C4).}
\label{figure1}
\end{figure}

Fig.~\ref{figure1} shows the DE of the EEG increments for the somnographic
channels O1,O2,C3 and C4 of a single individual. We see how for each channel
the DE: 1) reaches a saturation level for each channel, 2) has an ``alpha'' (%
$\sim $7.6 HZ in the case of this individual) modulation which is attenuated
with time, and 3) has a small amplitude residual asymptotic modulation. The
early-time modulation, with variable frequency in the alpha range and
variable amplitude, is observed in the somnographic channels for all
subjects. The saturation effect is present in all channels for all subjects
and it should be pointed out that this saturation is neither a consequence
of the finite length of the time series, nor of the finite amplitude of the
EEG signal. In fact if the data points were randomly rearranged, thereby
destroying any long time correlation in the time series, the EEG entropy
does not saturate. Consequently, this saturation effect is due to brain
dynamics and is not an artifact of the data processing. Robinson \cite
{robinson} observed this saturation in the calculation of the EEG second
moment and interpreted it as being due to dendritic filtering. The inset in
Fig.~\ref{figure1} depicts the \textit{pdf}s $p_{\text{sat}}(x)$, after the
entropy saturation is attained. These distributions have approximatively
exponential tails. 
% and a width that changes periodically in time as was first
%noted without further comment by Schl\"{o}gl \textit{et al}. \cite{schlogl99}%
%%%%%%%%%%%%%%%%%%%%%%%%%%Figure1

\section{EEG model}
The simplest dynamic model, which includes fluctuations, modulation
and dissipation, in short, all the properties displayed in Fig.~\ref{figure1}, 
has the form of a Langevin equation. We assume a dissipative linear
dynamic process $X(t)$, i.e., an Ornstein-Uhlenbeck process, with a periodic
driver having a random amplitude and frequency and an additive random force $%
\eta \left( t\right) $ which is a delta correlated Gaussian process of
strength $D$: 
\begin{equation}
\frac{dX(t)}{dt}=-\lambda X(t)+\eta \left( t\right) +\underset{j=0}{\sum }%
A_{j}\chi \left[ I_{j,s}\right] \sin \left[ 2\pi f_{j}t\right]
\label{sinusoidal}
\end{equation}
The coefficient $\lambda$$>$$0$ defines a negative
feedback, $\chi$$\left[I_{j,s}\right]$$=$1 when its argument is in the time
interval $I_{j,s}$$=$$[jt_{s},(j+1)t_{s}]$ and is zero otherwise, and $t_{s}$
is the 'stability' time after which a new frequency $f_{j}$ and a new
amplitude $A_{j}$ are selected.

The values of the frequencies $f_{j}$ and amplitudes $A_{j}$ are calculated
as follow. First, we calculate the spectral density in the time-frequency
domain of time series of EEG increments with a time resolution $t_{s}$ and a
frequency resolution $\Delta f$ by means of a Windowed Fourier Transform.
The spectral density, called the spectrogram (e.g. \cite{mallat}), is a
three-dimensional plot of the spectrum of the EEG increments $\xi _{j}$ as
it changes over time. 
%The spectrogram is the spectrum of the time series for
%a given time resolution, but which changes as a function of time, and
%therefore there is no one spectrum to characterize the process as we sweep
%through the non-stationary time series. 
Then, for each time interval of duration $t_{s}$ we consider the range of
frequencies of the alpha waves, 7-12 Hz, and find which frequency has the
maximum amplitude in the spectrogram. This procedure defines the frequency
and the amplitude of the time interval considered.

\begin{figure}[tbp]
\includegraphics[angle=0,width=1.0\linewidth,height=9cm]{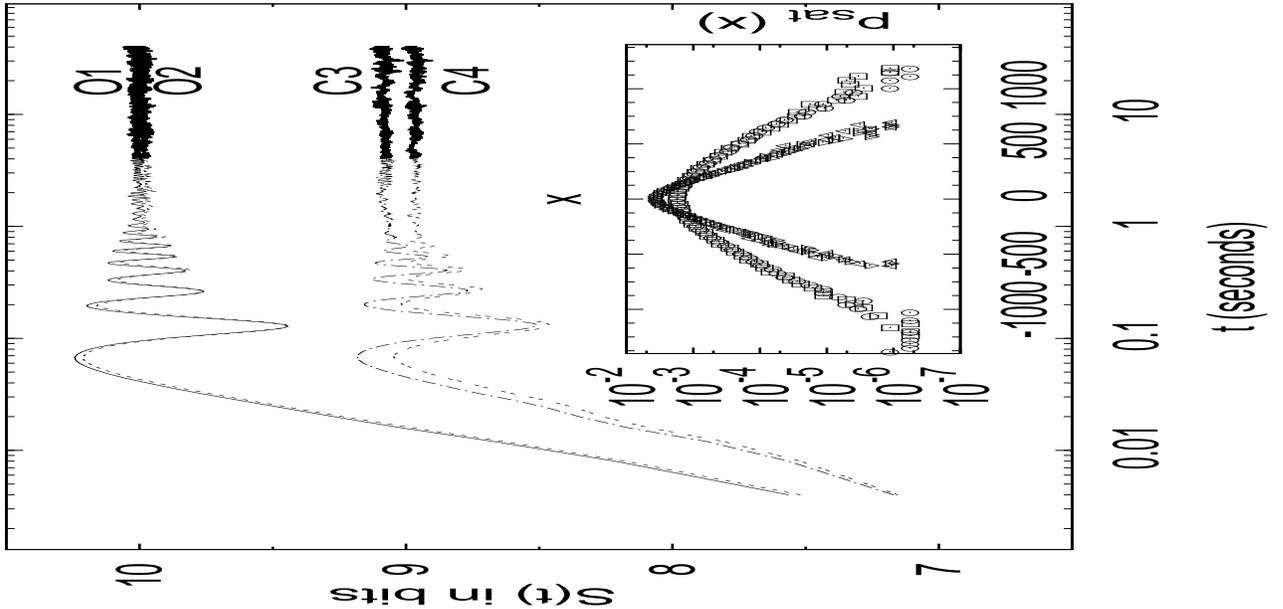}
\caption{(a) Spectrogram of the increments of channel O1. We plot the
base-10 logarithm of the spectral density. The time resolution is $t_{s}$$=$%
0.5 s, the frequency resolution is $\Delta f$$=$0.5 Hz. (b) Sequence of the
maxima of the spectrogram amplitude (normalized so that the maximum
amplitude is 1). This sequence is the sequence of coefficients $A_{j}$ used
in Eq.~( \ref{sinusoidal}). (c) Sequence of the frequencies corresponding to
the amplitude maxima of the spectrogram. This sequence is the sequence of
coefficient $f_{j}$ used in Eq.~(\ref{sinusoidal}).}
\label{figure2}
\end{figure}

Panel (a) of Fig.~\ref{figure2} shows the spectrogram relative to the
increments $\xi _{j}$ of the channel O1 for the same subject as in Fig.~\ref
{figure1}. Panels (b) and (c) of Fig.~\ref{figure2} show respectively the
sequence of amplitudes $A_{j}$ (normalized to a maximum amplitude of 1) and
of frequencies $f_{j}$ calculated using the procedure described above.
Without an \textit{a priori} knowledge of the typical duration of an alpha
wave packet, we set the stability time $t_{s}$ of Eq.~(\ref{sinusoidal})
equal to 0.5s. A time resolution of 0.5s and a frequency resolution of $\sim 
$ 0.5Hz in the spectrogram represent a reasonable time-frequency
localization for our purposes. %%%%%%%%%%%%%%%%%%%%%%%%%%Figure2

Consider the model case where $A_{j}$$=$$0$, for all $j$, no modulation is
present, and Eq.~(\ref{sinusoidal}) is the Ornstein-Uhlenbeck Langevin
equation. In this case the standard deviation of the variable $X$ is $\sigma
(t)$$=$ $\sqrt{D\left( 1-e^{-2\lambda t}\right) /\lambda }$. Consequently,
for $t$$\ll $1/$\lambda $ the entropy increases as $S(t)$$=$$C^{\prime }$$+$$%
\frac{1}{2}\log _{2}t$, with $C^{\prime }$$=$$C$$+$$\log _{2}$ $\sqrt{\frac{%
2D}{\lambda }}$ and a linear-log plot yields a straight line of slope $%
\delta $$=$$0.5$. For $t$$\gg $1/$\lambda $ the entropy reaches the
saturation level $S(t)$$=$$C$$+$$\log _{2}$$\sqrt{\frac{D}{\lambda }}$,
yielding an entropy structure similar to that of the EEG data depicted in
Fig.~\ref{figure1} without the modulation being present.

\begin{figure}[h]
\includegraphics[angle=-90,width=1.0\linewidth]{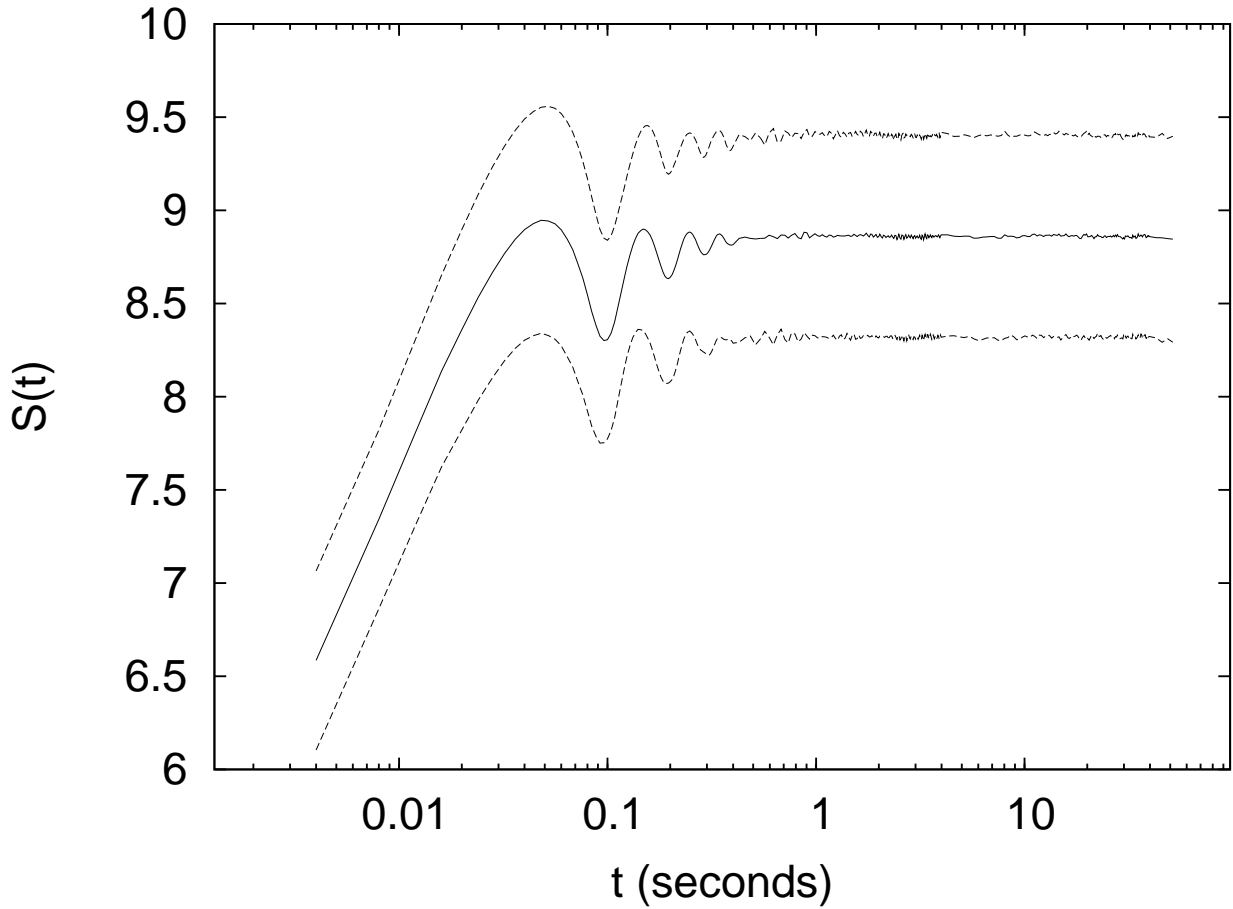}
\caption{Comparison between the diffusion entropy of the increments, solid
lines, of channel O1 and C3, and diffusion entropy of the increments,
points, of the varaible $X$ of Eq.~(\ref{sinusoidal}). The parameters used
in Eq.~(\ref{sinusoidal}) are $\protect\lambda $$=$0.055 D$=$40 for O1 and $%
\protect\lambda $$=$0.055 D$=$20 for C3. Inset show the comparison between
the pdfs at saturarion $\text{p}_{\text{sat}}(X)$$=$$p(X,t=2000)$: channels
O1 and C3, solid lines, variable $X$ of Eq.~(\ref{sinusoidal}), squares.}
\label{figure3}
\end{figure}

When the modulation is present $A_{j}$$\neq$$0$, Eq.~(\ref{sinusoidal}) is
numerically integrated, and the increments of the dynamic variable $X$ are
processed using the DE algorithm. In Fig.~\ref{figure3}, we compare the DE
obtained via Eq.~(\ref{sinusoidal}) with that of the channels O1 and C3,
already show in Fig.~\ref{figure1}. It is evident that the entropy
constructed from the solution to Eq.~(\ref{sinusoidal}) captures the
qualitative and many of the quantitative features of the DE of the EEG
increments. Moreover, the asymptotic \textit{pdf}s recorded in the inset
also agree with the empirical ones depicted in Fig.~\ref{figure1}. 
%%%%%%%%%%%%%%%%%%%%%%%Figure 3
%
In Table~1 we average the phenomenological parameters $\lambda $ and $D$ for
the somnographic channels for the 20 subjects in this study. 
\begin{table}[h]
\caption{The average values (avg.) and the standard deviations (s.d.) of the
parameters $\protect\lambda $ and $D$ of Eq.~(\ref{sinusoidal}) for all 20
subjects in this study.}
\label{table1}%\resizebox{!}{4.2cm}{
\begin{tabular}{|c|c|c|}
\hline
EEG channel & $\lambda$ (avg.$\pm$s.d.) & D (avg.$\pm$s.d.) \\ \hline
O1 & 0.0461$\pm$0.0187 & 16.37$\pm$6.88 \\ \hline
O2 & 0.0497$\pm$0.0182 & 16.35$\pm$6.72 \\ \hline
C3 & 0.0362$\pm$0.0186 & 10.19$\pm$3.90 \\ \hline
C4 & 0.0393$\pm$0.0200 & 10.60$\pm$3.72 \\ \hline
\end{tabular}
%}
\end{table}

\section{Conclusions}
The first notable property of the Langevin model is that it reaches a
saturation level, indicating that the EEG signal asymptotically carries a
maximum amount of information. The EEG entropy does not grow indefinitely as
would a random process with long-time correlation.Consequently, the EEG time
series do not simply scale as had been previously assumed by a number of
investigators \cite{scaling}.

The second notable property of the Langevin model is related to the first
and is the dissipation, or negative feedback, produced locally within the
channel of interest. The fluctuation-dissipation relation of Einstein is
what produces the maximum level of the entropy in a closed physical network,
and is given by the ratio of the strength of the additive fluctuations to
the dissipation rate. In the more general Langevin equation given here we do
not expect the saturation level to be given by this ratio alone, but to
depend on the asymptotic value of the 'variance' $\sigma \left( \infty
\right) $. Note that the asymptotic 'variance' may not be independent of
time, but contains residual information in the form of low amplitude beats
because of its dependence on the random near-periodic driver. This mechanism
also explains the saturation observed earlier \cite{robinson} by associating
the negative feedback with the dedritic filtering of the signal.

The third notable feature of the Langevin model is the attenuated
oscillation of the entropy in time. The attenuation occurs in the EEG
entropy because the alpha rhythm is not being generated at one source, but
is described as a collective property of the brain in that it is being
generated at a number of different locations \cite{basar97}. Here the
influence of the distributed sources is modeled by wave packets that persist
for a stability time $t_{s}$; one packet is replaced by another with a
slightly different carrier frequency and amplitude chosen from the empirical
spectrogram every time interval of length $t_{s}$. The concatenation of
these wave packets with fluctuating frequencies and amplitudes produces a
decoherence that attenuates the modulation of the resulting EEG entropy in
time. We show in the sequel that an average over a Gibbs ensemble of phase
fluctuations of a harmonic driver in a linear dissipative Langevin equation
can be solved analytically to obtain the variance $\sigma \left( t\right)
^{2}=D\left( 1-e^{-2\lambda t}\right) /\lambda +Ce^{-\gamma t}\left( 1-\cos
\Omega t\right) $. The constants $C$ and $\gamma $ are determined by the
distribution of amplitudes and frequencies given by the spectrogram, and $%
\Omega$ is the frequncy of the oscillation observe in the DE plot. This
analytic expression reproduces the resulsts of DE inllustrated in Fig.~\ref
{figure1}. Consequently, this modulation of the EEG entropy indicates that
the channel is coupled coherently to the rest of the brain. 
%numericaly integrating Eq.(4). 

The presence of the alpha rhythm modulation masks \cite{bimbini} any
early-time scaling property of the EEG dynamics. Eq.~(\ref{sinusoidal}) is
the simplest form of a fluctuation-dissipation process that implies the
presence of internal feedback to prevent the occurrence of large excursions
of the electric potential inside the brain. The presence of this negative
feedback mechanism cast doubts on the possibility of considering an EEG
record as the sum of two independent components, noise and trend, which is
the usual assumption made for the DFA method.

Finally, the analysis presented in this Letter support the notion that alpha
rhythms \cite{basar97}: 1) are not generated by one but by a number of
spatially distributed sources, whose relative incoherence produces the
attenuated modulation of the EEG entropy; 2) are not merely noise but are
random and near-periodic; and 3) do not represent passive states, but may
deteremine how different parts of the brain communicate.

\begin{acknowledgements}
The authors thank the Army Research Office for support of this research. The
code of the programs used for the EEG analysis (DE and spectrogram) is
available at \newline
http://www.duke.edu/$\sim$mi8/softwaresubpage/C$+$$+$\_programs.html
\end{acknowledgements}

%\begin{acknowledgements}
%\end{acknowledgements}

% BibTeX users please use
%\bibliographystyle{spmpsci}


\end{document}